# Basic Classes of Grammars with Prohibition

**Mark Burgin**

*University of California, Los Angeles*
Los Angeles, CA 90095, USA

**ABSTRACT**

A practical tool for natural language modeling and development of human-machine interaction is developed in the context of formal grammars and languages. A new type of formal grammars, called grammars with prohibition, is introduced. Grammars with prohibition provide more powerful tools for natural language generation and better describe processes of language learning than the conventional formal grammars. Here we study relations between languages generated by different grammars with prohibition based on conventional types of formal grammars such as context-free or context sensitive grammars. Besides, we compare languages generated by different grammars with prohibition and languages generated by conventional formal grammars. In particular, it is demonstrated that they have essentially higher computational power and expressive possibilities in comparison with the conventional formal grammars. Thus, while conventional formal grammars are recursive and subrecursive algorithms, many classes of grammars with prohibition are superrecursive algorithms. Results presented in this work are aimed at the development of human-machine interaction, modeling natural languages, empowerment of programming languages, computer simulation, better software systems, and theory of recursion.

## 1. Introduction

An important problem of computer technology is organization of convenient, flexible and efficient interaction with computers. It is important for many types of software systems in different areas: computer simulation, learning, decision-making, etc. Natural language is a tool for human-machine interaction that has several desirable properties. First, it provides an immediate vocabulary for talking about the contents of the computer. Second, it gives means of accessing information in the computer independently of its structure and encoding. Third, it shields the user from the formal access language of the underlying system. Fourth, it is available with a minimum of training. This is especially important for business and industry where natural language is the most preferable. As a result natural language comprehension and modeling is one of the central problems in artificial intelligence. Researchers have developed a quantity of different techniques to solve this problem.

Formal grammars were introduced by Chomsky (1956) in his paper on the syntactic structure of a natural language to the goal of representing natural languages by formal structures. In verbal communication, an utterance is characterized by the surface manifestation of a "deeper" structure representing "meaning" of the utterance. The deep structure can undergo a variety of transformations of form (e.g., changes of the word order, of endings, etc.) on its way up, while retaining its essential meaning. These transformations are performed by transformational grammars, which work with syntax. They have three components. The first component is a phrase-structure grammar generating strings of morphemes representing simple, declarative, active sentences, each with an associated phrase marker or derivation tree. The second component is a set of transformational rules for rearranging these strings and adding or deleting morphemes to form correct representations of the full variety of authorized sentences. Finally, a sequence of morphophonemic rules maps each sentence representation to a string of phonemes. Formal grammars are capable of describing much of the grammar, or syntax, of natural languages such as English or Spanish (Martin, 1991).

Later formal grammars were used to describe programming languages and build compilers. In this area, formal grammars became even more useful than in the province of natural languages. For instance, most of the syntax of such popular programming language as Pascal is described by Backus-Naur forms (Backus, 1959), which are equivalent to context-free grammars. Thus, formal grammars have played a central role in compiler technology and parser design since 1960's. More recently, these grammars have been intensively used to describe document formats for information exchange on the Web.

Formal grammars proved to be very efficient for generating various linguistic structures, but only for modeling small fragments of natural languages. Their generative and expressive power appeared insufficient for large linguistic systems, not speaking about such developed natural languages as English or Spanish. As Martin (1991) writes, it is unrealistic to expect to arrive at a complete description of natural languages using these grammars. As a result, the principal limitation of existing programs that perform natural language generation is that they fail to realize a sufficiently broad range of requirements to demonstrate convincing linguistic capability (Jacobs, 1986). All this brings us to the problem of achieving higher efficiency for formal grammars.

In this work, we further develop a new approach to this problem based on formal grammars with prohibition introduced and studied in (Burgin, 2005a; 2005b). Here we study relations between languages generated by different grammars with prohibition based on conventional types of grammars such as context-free or context sensitive grammars. Besides, we compare languages generated by different grammars with prohibition and languages generated by conventional formal grammars. In particular, it is demonstrated (cf., for example, Theorems 4, 6 and Corollary 2) that they have essentially higher computational power and expressive possibilities in comparison with the conventional formal grammars. As a result, they provide more means for human-machine interaction, modeling natural languages, empowerment of programming languages, computer simulation, developing better software, and theory of recursion.

The obtained results are summarized in the tables given in the Appendix, which represent relations between classes of languages generated by grammars with prohibition, as well as between languages generated by different grammars with prohibition and languages generated by conventional formal grammars.

It is necessary to remark that grammars with prohibition.were also studied by Carlucci, Case and Jain (2007), who called them correction grammars and used for learning in the limit of classes of recursively enumerable languages. Case and Jain (2011) proved the Rice and Rice-Shapiro theorems for transfinite correction grammars.

## 2. Grammars with Prohibition

To define formal grammars with prohibition, we fix some alphabet $\Sigma$ and consider languages and formal grammars that use only this alphabet.

**Definition 1.** A formal grammar $G$ with prohibition consists of rules that are divided into two parts: positive P$G$ and negative N$G$.

These rules generate in a conventional manner, i.e., by derivation or recursive inference (cf., for example, (Hopcroft et al, 2001)), two languages L(P$G$) and L(N$G$).

**Remark 1.** It is usually assumed that alphabet $\Sigma$ and systems of rules are finite.

**Definition 2.** We define the language of the grammar $G$ with prohibition as L($G$) = L(P$G$) \ L(N$G$).

Positive rules are used for generation (acceptation) words from the language, while negative rules are used for exclusion of incorrect forms.

**Remark 2.** When there are no negative rules, we obtain conventional formal grammars and their languages.

Construction of languages by means of grammars with prohibition correlates with the technique used by people for natural language text generation. At first, general rules for generating words and texts are given. Then exclusions from these general rules are described. Such exclusions mean prohibition of application of definite general rules in some cases. For instance, one of the simplest forms of a basic English sentence is

<subject> <verb> <object>

which is illustrated by the example

Sam wears a shirt.

However, there is a prohibition to use

A shirt wears Sam.

In some cases, it is possible to give all possible kinds of permitted sentences by positive rules. Yet often this becomes inefficient and it is more potent not to give all cases when a general rule may be applied, but to present those instances when application of the rule is prohibited. The same is true for generation of words. Irregular verbs give an example of such situation. Verbs in English and in many other languages come in two groups. Regular verbs such as "adopt", "behave", and "call" form their simple past tense and its past participle forms by adding the inflectional ending *-ed* (or in some cases *-d* or *-t*); this means that the past tense and the past participle of regular verbs are always identical in form. English has thousand of existing regular verbs, and new ones are being added all the time. The number of the irregular verbs is much smaller. About 180 verbs are irregular in standard English, and there have not been any recent new ones. In contrast to the regular verbs, past forms of the irregular verbs are unpredictable and demand remembering. Nevertheless, they have some patterns such as: "keep, kept", sleep, slept", "feel, felt", and "dream, dreamed"; "wear, wore", "bear, bore", "tear, tore", and "swear, swore"; "string, strung", "swing, swung", "sting, stung", and "fling, flung".

As the number of the irregular verbs is much smaller than the number of the regular verbs, it is much more efficient to keep in mind exclusion (or prohibition) rules for irregular verbs than to remember all regular verbs. In a formal way, at first all regular forms are generated for all verbs. Then these forms for irregular verbs are excluded from the language by negative rules. After this specific rules for irregular verbs fill the gap.

Construction of languages by means of grammars with prohibition is also adequate to learning processes. When an individual, a child or adult, learns some natural language, she/he receives information not only what is possible to do with words, but also what operations and constructions are forbidden. This situation is partially reflected in the general learning theory by the concept of co-learning (cf., for example, (Freivalds et al, 1994)) and learning with positive and negative examples. Procedures of co-learning are described by such grammars with prohibition in which positive rules generate the set of all words in the given alphabet, while negative rules allow one in an inductive mode (Burgin, 2003) to get the solution of the problem, i.e., to learn a given computable function.

Here, we consider classes of grammars with prohibition related to the Chomsky hierarchy (Chomsky, 1956; 1959).

### 3. Chomsky hierarchy of grammars and languages

The Chomsky hierarchy consists of the following levels:

1. Type-0 grammars (unrestricted or phrase structure grammars) include all conventional formal grammars and generate recursively enumerable languages, i.e., languages that are accepted by a Turing machine. We denote the class of unrestricted grammars by $G_0$ and the class of corresponding languages by $\mathbf{L}(G_0)$, i.e., of languages generated (computed or recognized) by grammars from $G_0$.

2. Type-1 grammars (context-sensitive grammars) generate the context-sensitive languages, which are exactly all languages that are accepted by a non-deterministic Turing machine whose tape is bounded by a constant times the length of the input. We denote the class of context-sensitive grammars by $G_1$ and the class of corresponding languages by $\mathbf{L}(G_1)$.

3. Type-2 grammars (context-free grammars) generate the context-free languages, which are exactly all languages that are accepted by a non-deterministic pushdown automaton. Context free languages are the theoretical basis for the syntax of most programming languages. We denote the class of context-free grammars by $G_2$ and the class of corresponding languages by $\mathbf{L}(G_2)$.

4. Type-3 grammars (regular grammars) generate the regular languages, which are exactly all languages that can be decided by a finite state automaton. Additionally, this family of formal languages can be obtained by regular expressions. Regular languages are commonly used to define search patterns and the lexical structure of programming languages. We denote the class of regular grammars by $G_3$ and the class of corresponding languages by $\mathbf{L}(G_3)$.

Every regular language is context-free, every context-free language is context-sensitive and every context-sensitive language is recursively enumerable. All inclusions are proper.

## 4. Grammars with prohibition related to Chomsky hierarchy

The class of grammars with prohibition in which the poitive grammar belongs to the class $G_i$ and the negaitive grammar belongs to the class $G_j$ is denoted by $G_{ij}$, while the class of corresponding languages, i.e., languages generated (computed or recognized) by grammars from $G_{ij}$, is denoted by $\mathbf{L}(G_{ij})$.

Thus, four types of conventional formal grammars give us 16 types of formal grammars with prohibition: $G_{00}, G_{01}, G_{02}, G_{03}, G_{10}, G_{11}, G_{12}, G_{13}, G_{20}, G_{21}, G_{22}, G_{23}, G_{30}, G_{31}, G_{32}, G_{33}$. This gives us 16 classes of formal languages: $\mathbf{L}(G_{00}), \mathbf{L}(G_{01}), \mathbf{L}(G_{02}), \mathbf{L}(G_{03}), \mathbf{L}(G_{10}), \mathbf{L}(G_{11}), \mathbf{L}(G_{12}), \mathbf{L}(G_{13}), \mathbf{L}(G_{20}), \mathbf{L}(G_{21}), \mathbf{L}(G_{22}), \mathbf{L}(G_{23}), \mathbf{L}(G_{30}), \mathbf{L}(G_{31}), \mathbf{L}(G_{32}), \mathbf{L}(G_{33})$. For instance, $\mathbf{L}(G_{03})$ consists of all formal languages that have the form $L_1 \setminus L_2$ where $L_1$ is an arbitrary recursively enumerable language and $L_2$ is an arbitrary regular language. A grammar $G$ that belongs to $G_{03}$ is called unrestricted\regular grammar and the corresponding language $L(G)$ is called enumerable\regular language. A grammar $G$ that belongs to $G_{12}$ is called context-sensitive\context-free grammar and the corresponding language $L(G)$ is called context-sensitive\context-free language. Our goal is to find relations between these classes.

**Theorem 1.** a) For all $i, j \in \{0, 1, 2, 3\}$, we have $\mathbf{L}(G_{ij}) \supseteq \mathbf{L}(G_i)$.

b) If $k > i$, then $\mathbf{L}(G_{ij}) \supseteq \mathbf{L}(G_{kj})$ and $\mathbf{L}(G_{ji}) \supseteq \mathbf{L}(G_{jk})$.

**Corollary 1**. For all $i \in \{0, 1, 2, 3\}$, we have $\mathbf{L}(G_{ii}) \supseteq \mathbf{L}(G_i)$.

Many of these inclusions are proper (cf., Theorem 7) but not all.

**Theorem 2.** $\mathbf{L}(G_{33}) = \mathbf{L}(G_3)$.

To describe and compare expresional power of grammars with prohibition, we use arithmetical hierarchy (Rogers, 1987). In it, the lowest level $\mathbf{\Sigma_0} = \mathbf{\Pi_0}$ consists of all recursively decidable (recursive) formal languages (sets). The next level has two parts: $\mathbf{\Sigma_1}$ consists of all recursively computable (recursively enumerable) formal languages (sets) and $\mathbf{\Pi_1}$ consists of all complements of recursively computable (recursively enumerable) formal languages (sets).

**Lemma 1.** If $L_D$ is a decidable and $L_E$ is an enumerable language, then $L = L_D \setminus L_E$ is a complement to an enumerable language.

Indeed, by properties of set-theoretical operations, $L = L_D \setminus L_E = \Sigma^* \setminus ((\Sigma^* \setminus L_D) \cup L_E )$. Then $L_1 = \Sigma^* \setminus L_D$ is a decidable language and the union of two enumerable languages is an enumerable language, i.e. $L_2 = (\Sigma^* \setminus L_D) \cup L_E$ is an enumerable language. Thus, $L = \Sigma^* \setminus L_2$ is a complement to the enumerable language $L_2$.

**Lemma 2.** If $L_D$ is a decidable and $L_E$ is an enumerable language, then $L = L_E \setminus L_D$ is an enumerable language.

Proof is similar to the proof of Lemma1.

**Theorem 3.** $\mathbf{L}(G_{03}) = \mathbf{\Sigma_1}$.

Proof is based on Lemma 2.

**Theorem 4.** $L(G_{30}) = \Pi_1$.

Proof is based on Lemma 1.

This result shows that in contrast to conventional formal grammars, formal grammars with prohibition can generate non-enumerable languages. Thus, the class $G_{30}$ and as we see below, $G_{20}$, $G_{10}$, and $G_{00}$ are classes of super-recursive algorithms (Burgin, 2005).

Theorems 1, 3, and 4 imply the following result.

**Corollary 2.** $L(G_{00}) = \Sigma_1 \cup \Pi_1$.

This result shows that formal grammars with prohibition have higher expressive (generative) power than conventional formal grammars and Turing machines. However, inductive Turing machines (Burgin, 2005) can compute or accept any language generated by a grammar with prohibition.

**Corollary 3.** $L(G_{00}) = L(G_{03}) \cup L(G_{30})$.

**Corollary 4.** $L(G_{01}) \cup L(G_{10}) = L(G_{02}) \cup L(G_{20}) = L(G_{03}) \cup L(G_{30})$.

**Theorem 5.** $L(G_{01}) = \Sigma_1$.

Proof is based on Lemma 2 as all context-sensitive languages are decidable.

Theorems 1 and 5 imply the following result.

**Corollary 5.** $L(G_{02}) = L(G_{01}) = L(G_{03}) = L(G_0) = \Sigma_1$.

**Theorem 6.** $L(G_{10}) = \Pi_1$.

Proof is based on Lemma 1 as all context-sensitive languages are decidable..

Theorems 1 and 6 imply the following result.

**Corollary 5.** $L(G_{20}) = L(G_{10}) = L(G_{30}) = \Pi_1$.

**Theorem 7.**  a) $L(G_{00}) \supset L(G_0)$, $L(G_{10}) \neq L(G_0)$, $L(G_{20}) \neq L(G_0)$ and $L(G_{30}) \neq L(G_0)$;

b) $L(G_{10}) \neq L(G_1)$, $L(G_{20}) \neq L(G_2)$, and $L(G_{30}) \neq L(G_3)$;

c) $L(G_{32}) \neq L(G_2)$, $L(G_{22}) \neq L(G_2)$, and $L(G_{12}) \neq L(G_2)$;

Indeed, inequalities and inclusions from parts a and b follow from previous results and relations between classes from the arithmetical hierarchy (Rogers, 1987). For c), we have

$L(G_{32}) \supset \Sigma^* \setminus L(G_2)$ and the class $L(G_2)$ of context-free languages is not closed under operation of difference.

At the same time, as the class $L(G_1)$ of context-sensitive languages is closed under operations of complement and intersection (Du and Ko, 2001), we have the following result.

**Theorem 8.** $L(G_{11}) = L(G_1)$.

**Theorem 9.** $L(G_{23}) = L(G_2)$.

Indeed, if $L_{CF}$ is a context-free and $L_R$ is a regular language, then $L = L_{CF} \setminus L_R = L_{CF} \cap (\Sigma^* \setminus L_R)$. Here $(\Sigma^* \setminus L_R)$ is a regular language and the class $L(G_2)$ of context-free languages is closed under operation of intersection with regular languages (Hopcroft, et al, 2001).

**Proposition 1.** $L(G_{32})$ is the complement of $L(G_2)$.

<u>Proof</u>. Let $L_{CF}$ be a context-free and $L_R$ be a regular language. For a subset $X$ of $\Sigma^*$, its complement is denoted by C$X$. Then $L = L_R \setminus L_{CF} = L_R \cap (\Sigma^* \setminus L_{CF}) = (\Sigma^* \setminus CL_R) \cap (\Sigma^* \setminus L_{CF}) = C(C(\Sigma^* \setminus CL_R) \cup C(\Sigma^* \setminus L_{CF})) = C(L_R \cup L_{CF}) = \Sigma^* \setminus L_1$ where $L_1$ is a context-free language because the class $L(G_2)$ of context-free languages is closed under operation of union (Du and Ko, 2001).

Proposition 1 is proved.

**Theorem 2.** $L(G_{32}) \neq L(G_1)$.

<u>Proof</u>. Let us assume that an arbitrary context-sensitive language $L_{CS}$ is equal to a complement C$L_{CF}$ of some context-free language $L_{CF}$. Then $L_{CF} = CL_{CS}$. However, C$L_{CS}$ is also a context-sensitive language as the class $L(G_1)$ of context-free languages is closed under operation of complement (Du and Ko, 2001). Moreover, as $L_{CS}$ is an arbitrary context-sensitive language, C$L_{CS}$ is also an arbitrary context-sensitive language. As there are context-sensitive languages that are not context-free, our assumption is false and theorem is proved.

**Conclusion**

We have considered grammars with prohibition that work with conventional data – strings of symbols or words – and generate traditional formal languages. Relations between classes of languages generated by grammars with prohibition obtained in this work, as well as relations between classes of languages generated by grammars with prohibition and classes of languages generated by conventional formal grammars are summarized in the tables from the Appendix.

However, grammars that work with more general objects than strings of symbols have been studied and found useful. For instance in (Murata, et al, 2001), grammars that work with trees are studied and applied to formal description of XML scheme languages. Formal grammars can work with arbitrary graphs and even with such complex objects as Kolmogorov complexes (Kolmogorov, 1953). Thus, it is interesting to investigate the following problem.

**Problem 1.** Consider grammars with prohibition that work with objects that are not strings and study their generative and expressive power.

An important peculiarity of formal grammars is that there is a correspondence between definite classes of grammars and types of abstract automata. For instance, regular grammars correspond to finite automata as they generate the class of languages. Context-free grammars correspond to pushdown automata, while unrestricted or phrase structure grammars correspond to Turing machines. This brings us to the following problem.

**Problem 2.** Develop correspondence between classes of grammars with prohibition and classes of automata.

When classes of languages are studied and used, it is useful to know their closure properties, i.e., with respect to what operations with languages they are closed and with respect to what operations with languages they are not closed. This brings us to the following problem.for grammars with prohibition.

**Problem 3.** Study closure properties of grammars with prohibition.

Besides, utilization of languages usually demands solving different algorithmic problems, e.g., whther the given language is empty or if the given word belong to the given language. This brings us to the following problem.for grammars with prohibition.

**Problem 4.** Study algorithmic problems of grammars with prohibition.

Here we considered only grammars with prohibition that correspond to the Chomsky hierarchy. However, there are many other types and kinds of formal grammars.

**Problem 5.** Study other types of grammars with prohibition, i.e., when positive and/or negative part of the grammar with prohibition does not belong to the Chomsky hierarchy.

For instance, the most noteworthy class of grammars lying properly between context-free and context-sensitive grammars is the class of indexed grammars (Aho, 1968; Parchmann and Duske, 1986). Consequently, the languages describable by indexed grammars - namely, the indexed languages - are a natural class of formal languages that form a proper superset of the context-free languages and a proper subset of the context-sensitive languages. Thus, we have the following problem.

**Problem 6.** Study grammars with prohibition when positive and/or negative part of the grammar with prohibition is an indexed grammar.

It is interesting to find, in particular whether the set of all indexed\indexed languages coincides with the set of all context-sensitive languages.

It would be also appealing to consider grammars and languages with prohibition when, at least, one of the grammars is a determistic context-free grammar (Hopcroft, et al, 2001).

Another popular class consists of programmed grammars. When a programmed grammar is used to derive a string, rule order is intrinsically predetermined by the availability of variables in the string under derivation. This process is generally non-deterministic because there may be several candidate rules. The idea of a programmed grammar is to impose an extrinsic ordering of rules reflecting a certain manner in which the generation process is envisaged by the composer. Thus, we have the following problem.

**Problem 7.** Study grammars with prohibition when positive and/or negative part of the grammar with prohibition is a programmed grammar.

An important class of formal grammars is formed by Boolean grammars and their generalizations (Okhotin, 2004). Thus, we have the following problem.

**Problem 8.** Study grammars with prohibition when positive and/or negative part of the grammar with prohibition is a Boolean grammar.

Tables in the Appendix, which represent relations between classes of languages generated by grammars with prohibition, leave two open problems.

**Problem 9.** Is the equality $\mathbf{L}(\boldsymbol{G}_{22}) = \mathbf{L}(\boldsymbol{G}_{11})$ true?

**Problem 10.** Is the equality $L(G_{22}) = L(G_1)$ true?

# Appendix

**Table 1.** Relations between languages of the grammars with prohibition

| type | 01 | 02 | 03 | 10 | 11 | 12 | 13 | 20 | 21 | 22 | 23 | 30 | 31 | 32 | 33 |
|------|----|----|----|----|----|----|----|----|----|----|----|----|----|----|----|
| 00 | ⊃ | ⊃ | ⊃ | ⊃ | ⊃ | ⊃ | ⊃ | ⊃ | ⊃ | ⊃ | ⊃ | ⊃ | ⊃ | ⊃ | ⊃ |
| 01 | = | = | = | ≠ | ⊃ | ⊃ | ⊃ | ≠ | ⊃ | ⊃ | ⊃ | ≠ | ⊃ | ⊃ | ⊃ |
| 02 | = | = | = | ≠ | ⊃ | ⊃ | ⊃ | ≠ | ⊃ | ⊃ | ⊃ | ≠ | ⊃ | ⊃ | ⊃ |
| 03 | = | = | = | ≠ | ⊃ | ⊃ | ⊃ | ≠ | ⊃ | ⊃ | ⊇ | ≠ | ⊃ | ⊃ | ⊃ |
| 10 | ≠ | ≠ | ≠ | = | ⊃ | ⊃ | ⊃ | = | ⊃ | ⊃ | ⊃ | = | ⊃ | ⊃ | ⊃ |
| 11 | ⊂ | ⊂ | ⊂ | ⊂ | = | = | = | ⊂ | = | ⊇ | ⊃ | ⊂ | = | ⊃ | ⊃ |
| 12 | ⊂ | ⊂ | ⊂ | ⊂ | = | = | = | ⊂ | = | ⊇ | ⊃ | ⊂ | = | ⊃ | ⊃ |
| 13 | ⊂ | ⊂ | ⊂ | ⊂ | = | = | = | ⊂ | = | ⊇ | ⊃ | ⊂ | = | ⊃ | ⊃ |
| 20 | ≠ | ≠ | ≠ | = | ⊃ | ⊃ | ⊃ | = | ⊃ | ⊃ | ⊃ | = | ⊃ | ⊃ | ⊃ |
| 21 | ⊂ | ⊂ | ⊂ | ⊂ | = | = | = | ⊂ | = | ⊇ | ⊃ | ⊂ | = | ⊃ | ⊃ |
| 22 | ⊂ | ⊂ | ⊂ | ⊂ | ⊆ | ⊆ | ⊆ | ⊂ | ⊆ | = | ⊃ | ⊂ | ⊃ | ⊃ | ⊃ |
| 23 | ⊂ | ⊂ | ⊂ | ⊂ | ⊂ | ⊂ | ⊂ | ⊂ | ⊂ | ⊂ | = | ⊂ | ⊂ | ≠ | ⊃ |
| 30 | ≠ | ≠ | ≠ | = | ⊃ | ⊃ | ⊃ | = | ⊃ | ⊃ | ⊃ | = | ⊃ | ⊃ | ⊃ |
| 31 | ⊂ | ⊂ | ⊂ | ⊂ | = | = | = | ⊂ | = | ⊃ | ⊃ | ⊂ | = | ⊃ | ⊃ |
| 32 | ⊂ | ⊂ | ⊂ | ⊂ | ⊂ | ⊂ | ⊂ | ⊂ | ⊂ | ⊂ | ≠ | ⊂ | ⊂ | = | ⊃ |
| 33 | ⊂ | ⊂ | ⊂ | ⊂ | ⊂ | ⊂ | ⊂ | ⊂ | ⊂ | ⊂ | ⊂ | ⊂ | ⊂ | ⊂ | = |

In this table, the pair *ij* means the class $\mathbf{L}(\boldsymbol{G}_{ij})$ of languages generated by the grammar $\boldsymbol{G}_{ij}$, that is by the grammar with prohibition in which the positive part is equal to $\boldsymbol{G}_i$ and the negative part is equal to $\boldsymbol{G}_j$. The symbol $\subset$ ($\subseteq$) in the row *ij* and column *kh* means that the class of languages $\mathbf{L}(\boldsymbol{G}_{ij})$ is included in (included in or equal to) the class of languages $\mathbf{L}(\boldsymbol{G}_{kh})$, while the symbol $\supset$ ($\supseteq$) in the row *ij* and column *kh* means that the class of languages $\mathbf{L}(\boldsymbol{G}_{kh})$ is included in (included in or equal to) the class of languages $\mathbf{L}(\boldsymbol{G}_{ij})$.

**Table 2.** Relations between languages of the grammars with prohibition and languages of the conventional formal grammars

| type | 00 | 01 | 02 | 03 | 10 | 11 | 12 | 13 | 20 | 21 | 22 | 23 | 30 | 31 | 32 | 33 |
|------|----|----|----|----|----|----|----|----|----|----|----|----|----|----|----|----|
| 0 | ⊂ | = | = | = | ≠ | ⊃ | ⊃ | ⊃ | ≠ | ⊃ | ⊃ | ⊃ | ≠ | ⊃ | ⊃ | ⊃ |
| 1 | ⊂ | ⊂ | ⊂ | ⊂ | ⊂ | = | = | = | ⊂ | = | ⊇ | ⊃ | ⊂ | = | ⊃ | ⊃ |
| 2 | ⊂ | ⊂ | ⊂ | ⊂ | ⊂ | ⊂ | ⊂ | ⊂ | ⊂ | ⊂ | ⊂ | = | ⊂ | ⊂ | ≠ | ⊃ |
| 3 | ⊂ | ⊂ | ⊂ | ⊂ | ⊂ | ⊂ | ⊂ | ⊂ | ⊂ | ⊂ | ⊂ | ⊂ | ⊂ | ⊂ | ⊂ | = |

In this table, the pair *ij* means the class $\mathbf{L}(\boldsymbol{G}_{ij})$ of languages generated by the grammar $\boldsymbol{G}_{ij}$ , that is by the grammar with prohibition in which the positive part is equal to $\boldsymbol{G}_i$ and the negative part is equal to $\boldsymbol{G}_j$.